\documentclass[reviewcopy]{elsarticle}

\usepackage[reviewcopy]{adndt}
\usepackage{longtable}

\usepackage{amsmath}

\usepackage{amssymb}

\usepackage{lscape}

\biboptions{square,sort&compress}
\bibpunct[]{[}{]}{,}{n}{}{;}
\begin{document}

\begin{frontmatter}

\journal{Atomic Data and Nuclear Data Tables}


\title{Nuclear magnetic shielding constants of Dirac one-electron atoms \\
in some low-lying discrete energy eigenstates}

  \author{Patrycja Stefa{\'n}ska\corref{cor1}}
  \ead{pstefanska@mif.pg.gda.pl}

  \cortext[cor1]{Corresponding author.}

  \address{Atomic and Optical Physics Division, Department of Atomic, Molecular and Optical Physics, \\
Faculty of Applied Physics and Mathematics, 
Gda{\'n}sk University of Technology, \\
Narutowicza 11/12, 80--233 Gda{\'n}sk, Poland}
 
\date{February 12, 2018}

\begin{abstract} 
We present tabulated data for the nuclear magnetic shielding constants ($\sigma$) of the Dirac one-electron atoms with a pointlike, motionless and spinless nucleus of charge $Ze$. Utilizing the exact general analytical formula for $\sigma$ derived by us \mbox{[P. Stefa{\'n}ska, Phys. Rev. A. 94 (2016) 012508/1-15],} valid for an arbitrary discrete energy eigenstate, we have computed the numerical values of the magnetic shielding factors for the ground state and for the first and the second set of excited states, i.e.: 2s$_{1/2}$, 2p$_{1/2}$, 2p$_{3/2}$, 3s$_{1/2}$, 3p$_{1/2}$, 3p$_{3/2}$, 3d$_{3/2}$, and 3d$_{5/2}$, of the relativistic hydrogenic ions with the nuclear charge numbers from the range $1 \leqslant Z \leqslant 137$. The comparisons of our results with the numerical values reported by other authors for some atomic states are also presented.
\end{abstract}

\end{frontmatter}

\textbf{Keywords:} Hydrogenlike atom, Shielding constant, Magnetic field, Screening factor, Nuclear magnetic resonance

\begin{center}
\large{\textbf{Published as: At. Data Nucl. Data Tables 120 (2018) 352--372}
\\*[1ex]
\textbf{doi: 10.1016/j.adt.2017.05.005}\\*[5ex]}
\end{center}




\newpage

\tableofcontents
\listofDtables
\listofDfigures
\vskip5pc


\section{Introduction}

When an atom or a molecule is placed in an external magnetic field, then electrons circling around their nuclei will begin to interact with the perturbing field. In this way, an additional magnetic field, oriented opposite to the external field, is generated in the system. Consequently, the strength of the magnetic field ``effectively sensed" by the nucleus decreases. This phenomenon is called nuclear magnetic shielding. The changes that occur in the location of the nucleus can be fully characterized by \emph{the magnetic shielding constant} (or \emph{the nuclear screening factor}). This physical quantity -- of course, depending on the electron density -- is directly related to the chemical shift, by which the position of the signal in the NMR (Nuclear Magnetic Resonance) spectrum is determined. This makes the magnetic shielding constant ($\sigma$) one of the most important atomic parameters currently used in chemistry and medicine.

There are many experimental results for the magnetic screening constants for many-electron atoms and molecules. But relativistic theoretical investigations of that quantity for such complex systems are quite complicated -- very often they come down to performing numerical calculations, which almost always require access to specialized software. The numerical values for $\sigma$ for some closed-shell atoms and ions were reported, among others, in Refs.\ \cite{Dalg62,Kolb82,John83}.  However, for the simplest systems, like one-electron atoms, the purely analytical relativistic calculations for the magnetic shielding constant (and also for another properties of the atom) are possible, and this was suggested already in 1969 by Hegstrom in Ref.\ \cite{Hegs69}. One of such a technique which allows one to derive analytically the formula for the shielding factor ($\sigma$) is the one based on the perturbation theory combined with the Sturmian expansion of the first-order generalized Dirac--Coulomb Green function, proposed in Ref.\ \cite{Szmy97}.  Using the method presented in that article, some time ago we have found the closed-form expression for $\sigma$ for the \emph{ground} state of relativistic hydrogenlike atom \cite{Stef12}, in agreement with corresponding formulas derived earlier by other authors in completely different ways. (For other applications of the technique proposed by Szmytkowski in Ref.\ \cite{Szmy97} to some electromagnetic properties of the ground state of Dirac one-electron atoms, see Refs.\ \cite{Szmy02b,Szmy04,Szmy02a,Miel06,Szmy11,Szmy12,Szmy14,Szmy16,Szmy16a}.).

Quite recently, we have shown that the usefulness of the aforementioned method goes beyond the study of the atomic ground state. In Refs.\ \cite{Stef15,Stef16a,Stef16b,Stef16,Stef17} we have considered an \emph{arbitrary} discrete energy eigenstate of relativistic hydrogenlike atom, in order to obtain analytical expressions (and further -- the numerical values) for some properties illustrating the influence of the external electromagnetic perturbations on the atom. In particular, in Ref.\ \cite{Stef16} one can find a detailed derivation of the formula for the magnetic shielding constant of Dirac one-electron atom being in an arbitrary energy eigenstate. The quantum state of the electron is generally characterized by the set of quantum numbers $\{n, \kappa, \mu \}$, in which $n$ denotes the radial quantum number, the Dirac quantum number $\kappa$ is an integer different form zero, whereas $\mu=-|\kappa|+\frac{1}{2}, -|\kappa|+\frac{3}{2}, \ldots, |\kappa|-\frac{1}{2}$ is the magnetic quantum number. The final closed-form expression for the shielding factor for such a state we have arrived at in Ref.\ \cite{Stef16} reads as follows:
\begin{eqnarray}
\sigma \equiv \sigma_{n\kappa \mu}=\frac{\alpha^2 Z}{N_{n\kappa}^2(4\kappa^2-1)} 
\Bigg[ 
\kappa^2-\frac{\eta_{n\kappa}^{(+)}}{4}-\frac{\eta_{n\kappa}^{(-)}}{4} + \frac{\mu^2}{4\kappa^2-1}\left(\frac{2\kappa+1}{2\kappa-1}\eta_{n\kappa}^{(+)}+\frac{2\kappa-1}{2\kappa+1}\eta_{n\kappa}^{(-)}\right)
\nonumber \\
 + \frac{4 \kappa^2 \mu^2}{4\kappa^2-1}
\left(
 \frac{4\kappa^2-5}{4\kappa^2-1}+ \frac{8\kappa[2\kappa(n+\gamma_{\kappa})-N_{n\kappa}](\alpha Z)^2}{\gamma_{\kappa}(4\gamma_{\kappa}^2-1)N_{n\kappa}^2}
\right)
\Bigg],
\label{eq:1}
\end{eqnarray}
where 
\begin{equation}
\eta_{n\kappa}^{(\pm)}=\frac{(2\kappa \pm 1)N_{n\kappa}}{n+\gamma_{\kappa}\pm N_{n\kappa}},
\label{eq:2}
\end{equation}
with
\begin{equation}
N_{n\kappa}=\sqrt{n^2+2n\gamma_{\kappa}+\kappa^2}
\label{eq:3}
\end{equation}
and
\begin{equation}
\gamma_{\kappa}=\sqrt{\kappa^2-(\alpha Z)^2}. 
\label{eq:4}
\end{equation}
In Ref.\ \cite{Stef16} we have proved analytically that the above result is valid for an arbitrary atomic state. However, in the aforementioned article, there are no tables with the numerical values of the relativistic shielding factors of hydrogenic ions. This kind of data may be useful -- among others -- to those who deal with the spectroscopic methods, for example, the NMR technique. This fact, as well as the lack of sufficiently large data sets in the literature, prompted us to carry out numerical calculations for the magnetic shielding constant for one-electron atoms ions with the nuclear charge numbers from the whole range $1 \leqslant Z \leqslant 137$. Our results will be discussed briefly in the next section.

\section{Discussion of results}

Numerical results presented in this work has been computed with the help of the exact analytical formula given in Eqs.\ (\ref{eq:1})--(\ref{eq:4}). In Table \ref{tab:1s_1-2} we have included the values of the magnetic shielding factors for the ground state (1s$_{1/2}$) of Dirac one-electron atoms. Tables \ref{tab:2s_1-2}--\ref{tab:2p_3-2_mu_3-2} contain the results for states belonging to the first set of excited state, i.e.: 2s$_{1/2}$, 2p$_{1/2}$, and 2p$_{3/2}$. The second excited atomic states (i.e.: 3s$_{1/2}$, 3p$_{1/2}$, 3p$_{3/2}$, 3d$_{3/2}$, and 3d$_{5/2}$) are presented in Tables \ref{tab:3s_1-2}--\ref{tab:3d_5-2_mu_5-2}. For the excited states, all possible values of the magnetic quantum number $\mu$ were considered. The Reader should observe that for the states with $\kappa=\pm 1$, there is a restriction for the nuclear charge number, i.e.: $Z<\alpha^{-1}\frac{\sqrt{3}}{2} \simeq 118.67$ (see Tables \ref{tab:1s_1-2}--\ref{tab:2p_1-2}, \ref{tab:3s_1-2}, and \ref{tab:3p_1-2}); a detailed explanation of this limitation can be found just follow Eq.\ (3.17) in Ref.\ \cite{Stef16}. 

The value of the inverse of the fine-structure constant we have used during creating Tables \ref{tab:1s_1-2}--\ref{tab:3d_5-2_mu_5-2} was $\alpha^{-1}=137.035 \: 999 \: 139$, and was taken from the newest CODATA 2014 report on Recommended Values of the Fundamental Physical Constants \cite{Mohr14}. However, in order to make the most accurate comparison of our exact results with the numerical results reported earlier by other authors for some atomic states \cite{Moor99,Pype99}, we have performed two additional tables with the values of $\sigma$ obtained using $\alpha^{-1}=137.035 \: 98 \:95$ (from CODATA 1986) \cite{Cohe87}. 

States with zero radial quantum number $n=0$ have been considered in Table \ref{tab:comparison}, where we have confronted our results with the values published by Moore in Ref.\ \cite{Moor99}. Actually, to be able to compare these numbers, we had to add the corresponding numerical components provided by her for a given atomic state. The resulting sums of the three ingredients for each state (the lower entries in Table \ref{tab:comparison}) appear to be in a pretty good agreement with our exact results (the upper entries therein).  

In contrast to the values of the screening constants presented in Tables \ref{tab:1s_1-2}--\ref{tab:comparison} (where they were given in units of $\alpha^2$), the numbers included in Table \ref{tab:comparison2} should be multiplied only by the commonly used factor of $10^{-6}$. We have rescaled in this way a part of the data from Tables \ref{tab:1s_1-2} and \ref{tab:2p_1-2}--\ref{tab:2p_3-2_mu_3-2} (i.e. for states 1s$_{1/2}$, 2p$_{1/2}$, and 2p$_{3/2}$), because we wanted to compare them with the corresponding numerical values provided by Pyper and Zhang \cite{Pype99}.  The agreement between their results and ours is also very good, and this has been shown in Table \ref{tab:comparison2}. The value of the magnetic shielding factor for the ground state of relativistic hydrogen atom ($Z=1$) can be found -- among others -- in the paper by Feiock and Johnson \cite{Feio69}. Their result, which is $\sigma=17.75 \times 10^{-6}$ (not shown in Table \ref{tab:comparison2}), also agrees with our result for this case.



\ack
I am indebted to Professor A.\ Rutkowski for his suggestion to publish present results. I also thank Professor R.\ Szmytkowski for technical assistance in carrying out the calculations.



\clearpage

\newpage

\TableExplanation

In all the tables we have used the following notation:

\bigskip
\renewcommand{\arraystretch}{1.0}

\begin{tabular}{ll}
$\sigma$ 			& Magnetic dipole shielding constant \\
$Z$ 					& Nuclear charge number \\
$\alpha$ 			& Fine structure constant \\
$\mu$         & Magnetic quantum number; $\mu=-j, \ldots, j$ \\*[1.5ex]
\multicolumn{2}{l}{The atomic states are referred to as $\mathcal{N}{x}_{j}$, according to the following guidelines:} \\*[1ex]
$\mathcal{N}$  & Principal quantum number \\
$\mathcal{N}=n+|\kappa|$ & Note that for the radial quantum number $n=0$, $\kappa$ is the negative integer! \\
$\kappa$ 			& Dirac quantum number; $\kappa=(l-j)(2j+1)$\\
$j$						& Total angular momentum quantum number; $j=|\kappa|-\frac{1}{2}=l \pm \frac{1}{2}$ \\
$x \equiv s$	& States with the orbital angular momentum quantum number $l=0$ \\
$x \equiv p$	& States with the orbital angular momentum quantum number  $l=1$ \\
$x \equiv d$	& States with the orbital angular momentum quantum number  $l=2$ \\
\end{tabular}

\datatables 



\setlength{\LTleft}{0pt}
\setlength{\LTright}{0pt} 


\setlength{\tabcolsep}{0.5\tabcolsep}

\renewcommand{\arraystretch}{1.0}

\footnotesize 
\begin{longtable}{@{\extracolsep\fill}rrrrrr@{}}
\caption{Magnetic shielding constants $\sigma$ (in units of $\alpha^2$) for the ground state (1s$_{1/2}$) of Dirac one-electron atoms. The number in brackets is the power of 10 by which the corresponding entry is to be multiplied. The value of the inverse of the fine-structure constant used was $\alpha^{-1}=137.035 \: 999 \: 139$ (from CODATA 2014).}
\label{tab:1s_1-2}
$Z$ & \multicolumn{1}{c}{$\sigma[\alpha^2]$} & $Z$ & \multicolumn{1}{c}{$\sigma[\alpha^2]$} & $Z$ & \multicolumn{1}{c}{$\sigma[\alpha^2]$} \\
\hline\\
\endfirsthead
\caption[]{(continued)}
$Z$ & \multicolumn{1}{c}{$\sigma[\alpha^2]$} & $Z$ & \multicolumn{1}{c}{$\sigma[\alpha^2]$} & $Z$ & \multicolumn{1}{c}{$\sigma[\alpha^2]$} \\
\hline\\
\endhead
1    &   3.333 811 647 331($-$1)   &   41   &   1.746 602 974 210($+$1)   &   81    &   7.817 869 668 184($+$1)    \\
2    &   6.670 494 089 568($-$1)   &   42   &   1.811 496 081 836($+$1)   &   82    &   8.174 063 737 041($+$1)    \\
3    &   1.001 292 267 930($+$0)    &   43   &   1.878 357 035 573($+$1)   &   83    &   8.552 930 762 593($+$1)    \\
4    &   1.336 398 189 894($+$0)    &   44   &   1.947 283 571 199($+$1)   &   84    &   8.956 539 142 795($+$1)    \\
5    &   1.672 656 996 173($+$0)    &   45   &   2.018 378 683 624($+$1)   &   85    &   9.387 209 193 087($+$1)    \\
6    &   2.010 360 345 643($+$0)    &   46   &   2.091 751 011 580($+$1)   &   86    &   9.847 552 267 330($+$1)    \\
7    &   2.349 802 205 046($+$0)    &   47   &   2.167 515 254 880($+$1)   &   87    &   1.034 051 735 170($+$2)    \\
8    &   2.691 279 326 667($+$0)    &   48   &   2.245 792 627 529($+$1)   &   88    &   1.086 944 684 530($+$2)    \\
\hline
\end{longtable}

\newpage

\footnotesize
\begin{longtable}{@{\extracolsep\fill}rrrrrr@{}}
\caption{Magnetic shielding constants $\sigma$ (in units of $\alpha^2$) of Dirac one-electron atoms in the excited state 2s$_{1/2}$. The number in brackets is the power of 10 by which the corresponding entry is to be multiplied. The value of the inverse of the fine-structure constant used was $\alpha^{-1}=137.035 \: 999 \: 139$ (from CODATA 2014).}
\label{tab:2s_1-2}
$Z$ & \multicolumn{1}{c}{$\sigma[\alpha^2]$} & $Z$ & \multicolumn{1}{c}{$\sigma[\alpha^2]$} & $Z$ & \multicolumn{1}{c}{$\sigma[\alpha^2]$} \\
\hline\\
\endfirsthead
\caption[]{(continued)}
$Z$ & \multicolumn{1}{c}{$\sigma[\alpha^2]$} & $Z$ & \multicolumn{1}{c}{$\sigma[\alpha^2]$} & $Z$ & \multicolumn{1}{c}{$\sigma[\alpha^2]$} \\
\hline\\
\endhead
1    &   8.334 039 107 064($-$2)   &   41   &   3.997 600 222 359($+$0)   &   81    &   1.571 351 017 454($+$1)    \\
2    &   1.667 231 453 748($-$1)   &   42   &   4.130 373 857 431($+$0)   &   82    &   1.641 605 092 176($+$1)    \\
3    &   2.501 907 102 596($-$1)   &   43   &   4.266 391 434 659($+$0)   &   83    &   1.716 707 475 740($+$1)    \\
4    &   3.337 857 016 011($-$1)   &   44   &   4.405 831 332 584($+$0)   &   84    &   1.797 140 369 287($+$1)    \\
5    &   4.175 509 888 277($-$1)   &   45   &   4.548 882 753 481($+$0)   &   85    &   1.883 447 735 330($+$1)    \\
6    &   5.015 297 808 566($-$1)   &   46   &   4.695 746 564 523($+$0)   &   86    &   1.976 245 167 495($+$1)    \\
7    &   5.857 657 132 582($-$1)   &   47   &   4.846 636 213 933($+$0)   &   87    &   2.076 231 677 664($+$1)    \\
8    &   6.703 029 369 684($-$1)   &   48   &   5.001 778 729 970($+$0)   &   88    &   2.184 203 844 590($+$1)    \\
\end{longtable}

\newpage

\footnotesize

\begin{longtable}{@{\extracolsep\fill}rrrrrr@{}}
\caption{Magnetic shielding constants $\sigma$ (in units of $\alpha^2$) of Dirac one-electron atoms in the excited state 2p$_{1/2}$. The number in brackets is the power of 10 by which the corresponding entry is to be multiplied. The value of the inverse of the fine-structure constant used was $\alpha^{-1}=137.035 \: 999 \: 139$ (from CODATA 2014).}
\label{tab:2p_1-2}
$Z$ & \multicolumn{1}{c}{$\sigma[\alpha^2]$} & $Z$ & \multicolumn{1}{c}{$\sigma[\alpha^2]$} & $Z$ & \multicolumn{1}{c}{$\sigma[\alpha^2]$} \\
\hline\\
\endfirsthead
\caption[]{(continued)}
$Z$ & \multicolumn{1}{c}{$\sigma[\alpha^2]$} & $Z$ & \multicolumn{1}{c}{$\sigma[\alpha^2]$} & $Z$ & \multicolumn{1}{c}{$\sigma[\alpha^2]$} \\
\hline\\
\endhead
1    &   2.782 118 900 586($+$3)   &   41   &   7.073 605 130 669($+$1)   &   81    &   4.268 987 601 923($+$1)    \\
2    &   1.391 156 693 555($+$3)   &   42   &   6.920 391 087 899($+$1)   &   82    &   4.253 568 075 818($+$1)    \\
3    &   9.275 458 812 054($+$2)   &   43   &   6.774 808 572 768($+$1)   &   83    &   4.240 611 438 939($+$1)    \\
4    &   6.957 729 597 111($+$2)   &   44   &   6.636 351 171 133($+$1)   &   84    &   4.230 218 908 270($+$1)    \\
5    &   5.567 352 419 207($+$2)   &   45   &   6.504 558 340 066($+$1)   &   85    &   4.222 510 517 120($+$1)    \\
6    &   4.640 651 768 925($+$2)   &   46   &   6.379 010 482 398($+$1)   &   86    &   4.217 627 714 061($+$1)    \\
7    &   3.978 909 669 943($+$2)   &   47   &   6.259 324 655 342($+$1)   &   87    &   4.215 736 488 486($+$1)    \\
8    &   3.482 767 222 004($+$2)   &   48   &   6.145 150 821 268($+$1)   &   88    &   4.217 031 142 058($+$1)    \\
\hline
\end{longtable}

\newpage

\footnotesize
\begin{longtable}{@{\extracolsep\fill}rrrrrr@{}}
\caption{Magnetic shielding constants $\sigma$ (in units of $\alpha^2$) of Dirac one-electron atoms in the excited state 2p$_{3/2}$ ($\mu=\pm 1/2$). The number in brackets is the power of 10 by which the corresponding entry is to be multiplied. The value of the inverse of the fine-structure constant used was $\alpha^{-1}=137.035 \: 999 \: 139$ (from CODATA 2014).}
\label{tab:2p_3-2_mu_1-2}
$Z$ & \multicolumn{1}{c}{$\sigma[\alpha^2]$} & $Z$ & \multicolumn{1}{c}{$\sigma[\alpha^2]$} & $Z$ & \multicolumn{1}{c}{$\sigma[\alpha^2]$} \\
\hline\\
\endfirsthead
\caption[]{(continued)}
$Z$ & \multicolumn{1}{c}{$\sigma[\alpha^2]$} & $Z$ & \multicolumn{1}{c}{$\sigma[\alpha^2]$} & $Z$ & \multicolumn{1}{c}{$\sigma[\alpha^2]$} \\
\hline\\
\endhead
1    &   $-$2.781 968 897 415($+$3)   &   47   &   $-$5.514 965 913 733($+$1)   &   93    &   $-$2.165 593 616 341($+$1)               \\
2    &   $-$1.390 856 668 182($+$3)   &   48   &   $-$5.382 867 430 870($+$1)   &   94    &   $-$2.124 047 243 592($+$1)               \\
3    &   $-$9.270 957 955 406($+$2)   &   49   &   $-$5.255 790 538 763($+$1)   &   95    &   $-$2.083 138 472 906($+$1)               \\
4    &   $-$6.951 727 565 491($+$2)   &   50   &   $-$5.133 432 930 687($+$1)   &   96    &   $-$2.042 845 685 127($+$1)               \\
5    &   $-$5.559 848 448 573($+$2)   &   51   &   $-$5.015 515 982 506($+$1)   &   97    &   $-$2.003 148 109 115($+$1)               \\
6    &   $-$4.631 644 902 110($+$2)   &   52   &   $-$4.901 782 474 997($+$1)   &   98    &   $-$1.964 025 777 215($+$1)               \\
7    &   $-$3.968 398 755 243($+$2)   &   53   &   $-$4.791 994 574 013($+$1)   &   99    &   $-$1.925 459 483 366($+$1)               \\
8    &   $-$3.470 750 911 465($+$2)   &   54   &   $-$4.685 932 035 049($+$1)   &   100   &   $-$1.887 430 743 694($+$1)               \\
\hline
\end{longtable}

\newpage

\footnotesize
\begin{longtable}{@{\extracolsep\fill}rrrrrr@{}}
\caption{Magnetic shielding constants $\sigma$ (in units of $\alpha^2$) of Dirac one-electron atoms in the excited state 2p$_{3/2}$ ($\mu=\pm 3/2$). The number in brackets is the power of 10 by which the corresponding entry is to be multiplied. The value of the inverse of the fine-structure constant used was $\alpha^{-1}=137.035 \: 999 \: 139$ (from CODATA 2014).}
\label{tab:2p_3-2_mu_3-2}
$Z$ & \multicolumn{1}{c}{$\sigma[\alpha^2]$} & $Z$ & \multicolumn{1}{c}{$\sigma[\alpha^2]$} & $Z$ & \multicolumn{1}{c}{$\sigma[\alpha^2]$} \\
\hline\\
\endfirsthead
\caption[]{(continued)}
$Z$ & \multicolumn{1}{c}{$\sigma[\alpha^2]$} & $Z$ & \multicolumn{1}{c}{$\sigma[\alpha^2]$} & $Z$ & \multicolumn{1}{c}{$\sigma[\alpha^2]$} \\
\hline\\
\endhead
1    &   1.000 028 534 293($-$1)   &   47   &   5.006 732 203 585($+$0)   &   93    &   1.194 851 329 368($+$1)   \\
2    &   2.000 228 284 951($-$1)   &   48   &   5.127 227 928 174($+$0)   &   94    &   1.214 396 435 565($+$1)   \\
3    &   3.000 770 521 391($-$1)   &   49   &   5.248 650 771 844($+$0)   &   95    &   1.234 202 117 158($+$1)   \\
4    &   4.001 826 619 163($-$1)   &   50   &   5.371 024 936 168($+$0)   &   96    &   1.254 274 039 325($+$1)   \\
5    &   5.003 568 113 104($-$1)   &   51   &   5.494 374 957 625($+$0)   &   97    &   1.274 617 999 639($+$1)   \\
6    &   6.006 166 750 591($-$1)   &   52   &   5.618 725 718 030($+$0)   &   98    &   1.295 239 932 296($+$1)   \\
7    &   7.009 794 544 924($-$1)   &   53   &   5.744 102 455 240($+$0)   &   99    &   1.316 145 912 505($+$1)   \\
8    &   8.014 623 828 881($-$1)   &   54   &   5.870 530 774 109($+$0)   &   100   &   1.337 342 161 028($+$1)   \\
\hline
\end{longtable}

\newpage

\footnotesize
\begin{longtable}{@{\extracolsep\fill}rrrrrr@{}}
\caption{Magnetic shielding constants $\sigma$ (in units of $\alpha^2$) of Dirac one-electron atoms in the excited state 3s$_{1/2}$. The number in brackets is the power of 10 by which the corresponding entry is to be multiplied. The value of the inverse of the fine-structure constant used was $\alpha^{-1}=137.035 \: 999 \: 139$ (from CODATA 2014).}
\label{tab:3s_1-2}
$Z$ & \multicolumn{1}{c}{$\sigma[\alpha^2]$} & $Z$ & \multicolumn{1}{c}{$\sigma[\alpha^2]$} & $Z$ & \multicolumn{1}{c}{$\sigma[\alpha^2]$} \\
\hline\\
\endfirsthead
\caption[]{(continued)}
$Z$ & \multicolumn{1}{c}{$\sigma[\alpha^2]$} & $Z$ & \multicolumn{1}{c}{$\sigma[\alpha^2]$} & $Z$ & \multicolumn{1}{c}{$\sigma[\alpha^2]$} \\
\hline\\
\endhead
1    &   3.703 923 474 808($-$2)   &   41   &   1.698 307 552 238($+$0)   &   81    &   5.725 125 585 478($+$0)    \\
2    &   7.409 166 080 451($-$2)   &   42   &   1.750 586 151 557($+$0)   &   82    &   5.948 718 328 129($+$0)    \\
3    &   1.111 704 947 074($-$1)   &   43   &   1.803 857 232 510($+$0)   &   83    &   6.186 797 501 693($+$0)    \\
4    &   1.482 890 035 448($-$1)   &   44   &   1.858 174 655 865($+$0)   &   84    &   6.440 797 647 070($+$0)    \\
5    &   1.854 605 304 549($-$1)   &   45   &   1.913 595 512 887($+$0)   &   85    &   6.712 336 790 020($+$0)    \\
6    &   2.226 985 204 001($-$1)   &   46   &   1.970 180 375 848($+$0)   &   86    &   7.003 245 748 591($+$0)    \\
7    &   2.600 165 463 054($-$1)   &   47   &   2.027 993 570 833($+$0)   &   87    &   7.315 603 132 682($+$0)    \\
8    &   2.974 283 356 588($-$1)   &   48   &   2.087 103 475 176($+$0)   &   88    &   7.651 777 353 963($+$0)    \\
\hline
\end{longtable}

\newpage

\footnotesize
\begin{longtable}{@{\extracolsep\fill}rrrrrr@{}}
\caption{Magnetic shielding constants $\sigma$ (in units of $\alpha^2$) of Dirac one-electron atoms in the excited state 3p$_{1/2}$. The number in brackets is the power of 10 by which the corresponding entry is to be multiplied. The value of the inverse of the fine-structure constant used was $\alpha^{-1}=137.035 \: 999 \: 139$ (from CODATA 2014).}
\label{tab:3p_1-2}
$Z$ & \multicolumn{1}{c}{$\sigma[\alpha^2]$} & $Z$ & \multicolumn{1}{c}{$\sigma[\alpha^2]$} & $Z$ & \multicolumn{1}{c}{$\sigma[\alpha^2]$} \\
\hline\\
\endfirsthead
\caption[]{(continued)}
$Z$ & \multicolumn{1}{c}{$\sigma[\alpha^2]$} & $Z$ & \multicolumn{1}{c}{$\sigma[\alpha^2]$} & $Z$ & \multicolumn{1}{c}{$\sigma[\alpha^2]$} \\
\hline\\
\endhead
1    &   2.782 082 890 470($+$3)   &   41   &   6.910 908 149 276($+$1)   &   81    &   3.768 455 123 575($+$1)    \\
2    &   1.391 084 662 008($+$3)   &   42   &   6.752 836 629 589($+$1)   &   82    &   3.735 948 908 009($+$1)    \\
3    &   9.274 378 055 773($+$2)   &   43   &   6.602 317 513 960($+$1)   &   83    &   3.704 927 797 438($+$1)    \\
4    &   6.956 288 059 806($+$2)   &   44   &   6.458 840 370 844($+$1)   &   84    &   3.675 401 178 243($+$1)    \\
5    &   5.565 549 646 388($+$2)   &   45   &   6.321 940 416 759($+$1)   &   85    &   3.647 385 858 656($+$1)    \\
6    &   4.638 487 191 071($+$2)   &   46   &   6.191 193 574 145($+$1)   &   86    &   3.620 906 878 910($+$1)    \\
7    &   3.976 382 601 559($+$2)   &   47   &   6.066 212 161 873($+$1)   &   87    &   3.595 998 504 140($+$1)    \\
8    &   3.479 876 860 479($+$2)   &   48   &   5.946 641 126 318($+$1)   &   88    &   3.572 705 440 162($+$1)    \\
\hline
\end{longtable}

\newpage

\footnotesize
\begin{longtable}{@{\extracolsep\fill}rrrrrr@{}}
\caption{Magnetic shielding constants $\sigma$ (in units of $\alpha^2$) of Dirac one-electron atoms in the excited state 3p$_{3/2}$ ($\mu=\pm 1/2$). The number in brackets is the power of 10 by which the corresponding entry is to be multiplied. The value of the inverse of the fine-structure constant used was $\alpha^{-1}=137.035 \: 999 \: 139$ (from CODATA 2014).}
\label{tab:3p_3-2_mu_1-2}
$Z$ & \multicolumn{1}{c}{$\sigma[\alpha^2]$} & $Z$ & \multicolumn{1}{c}{$\sigma[\alpha^2]$} & $Z$ & \multicolumn{1}{c}{$\sigma[\alpha^2]$} \\
\hline\\
\endfirsthead
\caption[]{(continued)}
$Z$ & \multicolumn{1}{c}{$\sigma[\alpha^2]$} & $Z$ & \multicolumn{1}{c}{$\sigma[\alpha^2]$} & $Z$ & \multicolumn{1}{c}{$\sigma[\alpha^2]$} \\
\hline\\
\endhead
1    &   $-$2.782 016 222 669($+$3)   &   47   &   $-$5.739 014 103 697($+$1)   &   93    &   $-$2.619 661 547 266($+$1)               \\
2    &   $-$1.390 951 319 594($+$3)   &   48   &   $-$5.611 756 338 335($+$1)   &   94    &   $-$2.583 348 334 573($+$1)               \\
3    &   $-$9.272 377 749 217($+$2)   &   49   &   $-$5.489 525 051 725($+$1)   &   95    &   $-$2.547 686 251 175($+$1)               \\
4    &   $-$6.953 620 666 150($+$2)   &   50   &   $-$5.372 018 064 047($+$1)   &   96    &   $-$2.512 653 963 211($+$1)               \\
5    &   $-$5.562 214 892 304($+$2)   &   51   &   $-$5.258 956 879 768($+$1)   &   97    &   $-$2.478 230 991 076($+$1)               \\
6    &   $-$4.634 484 734 216($+$2)   &   52   &   $-$5.150 084 410 021($+$1)   &   98    &   $-$2.444 397 665 079($+$1)               \\
7    &   $-$3.971 712 030 117($+$2)   &   53   &   $-$5.045 162 952 819($+$1)   &   99    &   $-$2.411 135 083 748($+$1)               \\
8    &   $-$3.474 537 692 607($+$2)   &   54   &   $-$4.943 972 397 676($+$1)   &   100   &   $-$2.378 425 074 608($+$1)               \\
\hline
\end{longtable}

\newpage

\footnotesize
\begin{longtable}{@{\extracolsep\fill}rrrrrr@{}}
\caption{Magnetic shielding constants $\sigma$ (in units of $\alpha^2$) of Dirac one-electron atoms in the excited state 3p$_{3/2}$ ($\mu=\pm 3/2$). The number in brackets is the power of 10 by which the corresponding entry is to be multiplied. The value of the inverse of the fine-structure constant used was $\alpha^{-1}=137.035 \: 999 \: 139$ (from CODATA 2014).}
\label{tab:3p_3-2_mu_3-2}
$Z$ & \multicolumn{1}{c}{$\sigma[\alpha^2]$} & $Z$ & \multicolumn{1}{c}{$\sigma[\alpha^2]$} & $Z$ & \multicolumn{1}{c}{$\sigma[\alpha^2]$} \\
\hline\\
\endfirsthead
\caption[]{(continued)}
$Z$ & \multicolumn{1}{c}{$\sigma[\alpha^2]$} & $Z$ & \multicolumn{1}{c}{$\sigma[\alpha^2]$} & $Z$ & \multicolumn{1}{c}{$\sigma[\alpha^2]$} \\
\hline\\
\endhead
1    &   4.444 542 008 200($-$2)   &   47   &   2.194 987 914 398($+$0)   &   93    &   5.083 238 724 003($+$0)   \\
2    &   8.889 669 447 279($-$2)   &   48   &   2.246 579 299 652($+$0)   &   94    &   5.163 010 060 919($+$0)   \\
3    &   1.333 596 798 990($-$1)   &   49   &   2.298 500 313 290($+$0)   &   95    &   5.243 811 083 108($+$0)   \\
4    &   1.778 402 379 248($-$1)   &   50   &   2.350 760 108 989($+$0)   &   96    &   5.325 667 295 712($+$0)   \\
5    &   2.223 442 373 750($-$1)   &   51   &   2.403 367 999 735($+$0)   &   97    &   5.408 604 900 878($+$0)   \\
6    &   2.668 775 567 636($-$1)   &   52   &   2.456 333 463 074($+$0)   &   98    &   5.492 650 821 928($+$0)   \\
7    &   3.114 460 867 302($-$1)   &   53   &   2.509 666 146 488($+$0)   &   99    &   5.577 832 728 494($+$0)   \\
8    &   3.560 557 324 843($-$1)   &   54   &   2.563 375 872 945($+$0)   &   100   &   5.664 179 062 647($+$0)   \\
\hline
\end{longtable}

\newpage

\footnotesize
\begin{longtable}{@{\extracolsep\fill}rrrrrr@{}}
\caption{Magnetic shielding constants $\sigma$ (in units of $\alpha^2$) of Dirac one-electron atoms in the excited state 3d$_{3/2}$ ($\mu=\pm 1/2$). The number in brackets is the power of 10 by which the corresponding entry is to be multiplied. The value of the inverse of the fine-structure constant used was $\alpha^{-1}=137.035 \: 999 \: 139$ (from CODATA 2014).}
\label{tab:3d_3-2_mu_1-2}
$Z$ & \multicolumn{1}{c}{$\sigma[\alpha^2]$} & $Z$ & \multicolumn{1}{c}{$\sigma[\alpha^2]$} & $Z$ & \multicolumn{1}{c}{$\sigma[\alpha^2]$} \\
\hline\\
\endfirsthead
\caption[]{(continued)}
$Z$ & \multicolumn{1}{c}{$\sigma[\alpha^2]$} & $Z$ & \multicolumn{1}{c}{$\sigma[\alpha^2]$} & $Z$ & \multicolumn{1}{c}{$\sigma[\alpha^2]$} \\
\hline\\
\endhead
1    &   1.802 795 342 177($+$3)   &   47   &   3.951 137 250 530($+$1)   &   93    &   2.175 154 669 680($+$1)               \\
2    &   9.014 341 164 174($+$2)   &   48   &   3.873 740 825 098($+$1)   &   94    &   2.157 344 556 779($+$1)               \\
3    &   6.009 965 739 946($+$2)   &   49   &   3.799 609 803 823($+$1)   &   95    &   2.139 978 768 223($+$1)               \\
4    &   4.507 899 541 760($+$2)   &   50   &   3.728 548 584 369($+$1)   &   96    &   2.123 043 947 913($+$1)               \\
5    &   3.606 757 053 822($+$2)   &   51   &   3.660 376 914 457($+$1)   &   97    &   2.106 527 303 938($+$1)               \\
6    &   3.006 076 442 268($+$2)   &   52   &   3.594 928 416 047($+$1)   &   98    &   2.090 416 580 185($+$1)               \\
7    &   2.577 088 353 792($+$2)   &   53   &   3.532 049 276 600($+$1)   &   99    &   2.074 700 029 684($+$1)               \\
8    &   2.255 408 115 306($+$2)   &   54   &   3.471 597 085 764($+$1)   &   100   &   2.059 366 389 564($+$1)               \\
\hline

\end{longtable}

\newpage

\footnotesize
\begin{longtable}{@{\extracolsep\fill}rrrrrr@{}}
\caption{Magnetic shielding constants $\sigma$ (in units of $\alpha^2$) of Dirac one-electron atoms in the excited state 3d$_{3/2}$ ($\mu=\pm 3/2$). The number in brackets is the power of 10 by which the corresponding entry is to be multiplied. The value of the inverse of the fine-structure constant used was $\alpha^{-1}=137.035 \: 999 \: 139$ (from CODATA 2014).}
\label{tab:3d_3-2_mu_3-2}
$Z$ & \multicolumn{1}{c}{$\sigma[\alpha^2]$} & $Z$ & \multicolumn{1}{c}{$\sigma[\alpha^2]$} & $Z$ & \multicolumn{1}{c}{$\sigma[\alpha^2]$} \\
\hline\\
\endfirsthead
\caption[]{(continued)}
$Z$ & \multicolumn{1}{c}{$\sigma[\alpha^2]$} & $Z$ & \multicolumn{1}{c}{$\sigma[\alpha^2]$} & $Z$ & \multicolumn{1}{c}{$\sigma[\alpha^2]$} \\
\hline\\
\endhead
1    &   1.201 888 253 354($+$3)   &   47   &   2.756 044 695 895($+$1)   &   93    &   1.732 057 778 713($+$1)   \\
2    &   6.010 054 646 848($+$2)   &   48   &   2.707 313 019 753($+$1)   &   94    &   1.724 582 439 587($+$1)   \\
3    &   4.007 384 714 441($+$2)   &   49   &   2.660 776 483 764($+$1)   &   95    &   1.717 458 526 708($+$1)   \\
4    &   3.006 254 364 150($+$2)   &   50   &   2.616 305 183 525($+$1)   &   96    &   1.710 678 455 982($+$1)   \\
5    &   2.405 739 951 535($+$2)   &   51   &   2.573 779 456 173($+$1)   &   97    &   1.704 235 054 005($+$1)   \\
6    &   2.005 533 620 160($+$2)   &   52   &   2.533 088 896 773($+$1)   &   98    &   1.698 121 540 307($+$1)   \\
7    &   1.719 789 167 354($+$2)   &   53   &   2.494 131 486 101($+$1)   &   99    &   1.692 331 510 790($+$1)   \\
8    &   1.505 583 510 670($+$2)   &   54   &   2.456 812 815 372($+$1)   &   100   &   1.686 858 922 295($+$1)   \\
\hline
\end{longtable}

\newpage

\footnotesize
\begin{longtable}{@{\extracolsep\fill}rrrrrr@{}}
\caption{Magnetic shielding constants $\sigma$ (in units of $\alpha^2$) of Dirac one-electron atoms in the excited state 3d$_{5/2}$ ($\mu=\pm 1/2$). The number in brackets is the power of 10 by which the corresponding entry is to be multiplied. The value of the inverse of the fine-structure constant used was $\alpha^{-1}=137.035 \: 999 \: 139$ (from CODATA 2014).}
\label{tab:3d_5-2_mu_1-2}
$Z$ & \multicolumn{1}{c}{$\sigma[\alpha^2]$} & $Z$ & \multicolumn{1}{c}{$\sigma[\alpha^2]$} & $Z$ & \multicolumn{1}{c}{$\sigma[\alpha^2]$} \\
\hline\\
\endfirsthead
\caption[]{(continued)}
$Z$ & \multicolumn{1}{c}{$\sigma[\alpha^2]$} & $Z$ & \multicolumn{1}{c}{$\sigma[\alpha^2]$} & $Z$ & \multicolumn{1}{c}{$\sigma[\alpha^2]$} \\
\hline\\
\endhead
1    &   $-$1.802 737 140 970($+$3)   &   47   &   $-$3.676 015 209 631($+$1)   &   93    &   $-$1.620 610 226 415($+$1)   \\
2    &   $-$9.013 177 131 173($+$2)   &   48   &   $-$3.592 693 915 741($+$1)   &   94    &   $-$1.596 512 048 885($+$1)   \\
3    &   $-$6.008 219 668 249($+$2)   &   49   &   $-$3.512 633 354 025($+$1)   &   95    &   $-$1.572 846 006 106($+$1)   \\
4    &   $-$4.505 571 404 725($+$2)   &   50   &   $-$3.435 637 805 826($+$1)   &   96    &   $-$1.549 598 501 840($+$1)   \\
5    &   $-$3.603 846 815 919($+$2)   &   51   &   $-$3.361 526 901 246($+$1)   &   97    &   $-$1.526 756 499 035($+$1)   \\
6    &   $-$3.002 584 059 068($+$2)   &   52   &   $-$3.290 134 143 292($+$1)   &   98    &   $-$1.504 307 491 265($+$1)   \\
7    &   $-$2.573 013 771 957($+$2)   &   53   &   $-$3.221 305 599 098($+$1)   &   99    &   $-$1.482 239 475 915($+$1)   \\
8    &   $-$2.250 751 272 575($+$2)   &   54   &   $-$3.154 898 736 566($+$1)   &   100   &   $-$1.460 540 928 970($+$1)   \\
\hline
\end{longtable}

\newpage

\footnotesize
\begin{longtable}{@{\extracolsep\fill}rrrrrr@{}}
\caption{Magnetic shielding constants $\sigma$ (in units of $\alpha^2$) of Dirac one-electron atoms in the excited state 3d$_{5/2}$ ($\mu=\pm 3/2$). The number in brackets is the power of 10 by which the corresponding entry is to be multiplied. The value of the inverse of the fine-structure constant used was $\alpha^{-1}=137.035 \: 999 \: 139$ (from CODATA 2014).}
\label{tab:3d_5-2_mu_3-2}
$Z$ & \multicolumn{1}{c}{$\sigma[\alpha^2]$} & $Z$ & \multicolumn{1}{c}{$\sigma[\alpha^2]$} & $Z$ & \multicolumn{1}{c}{$\sigma[\alpha^2]$} \\
\hline\\
\endfirsthead
\caption[]{(continued)}
$Z$ & \multicolumn{1}{c}{$\sigma[\alpha^2]$} & $Z$ & \multicolumn{1}{c}{$\sigma[\alpha^2]$} & $Z$ & \multicolumn{1}{c}{$\sigma[\alpha^2]$} \\
\hline\\
\endhead
1    &   $-$1.201 808 887 437($+$3)   &   47   &   $-$2.374 033 490 069($+$1)   &   93    &   $-$9.162 820 482 311($+$0)   \\
2    &   $-$6.008 467 278 303($+$2)   &   48   &   $-$2.316 764 285 622($+$1)   &   94    &   $-$8.980 691 814 721($+$0)   \\
3    &   $-$4.005 003 536 060($+$2)   &   49   &   $-$2.261 663 037 289($+$1)   &   95    &   $-$8.801 311 165 362($+$0)   \\
4    &   $-$3.003 079 225 213($+$2)   &   50   &   $-$2.208 599 131 022($+$1)   &   96    &   $-$8.624 585 901 323($+$0)   \\
5    &   $-$2.401 770 651 011($+$2)   &   51   &   $-$2.157 452 184 453($+$1)   &   97    &   $-$8.450 427 096 045($+$0)   \\
6    &   $-$2.000 769 906 643($+$2)   &   52   &   $-$2.108 111 062 974($+$1)   &   98    &   $-$8.278 749 338 578($+$0)   \\
7    &   $-$1.714 230 738 988($+$2)   &   53   &   $-$2.060 473 007 195($+$1)   &   99    &   $-$8.109 470 554 358($+$0)   \\
8    &   $-$1.499 230 015 052($+$2)   &   54   &   $-$2.014 442 857 353($+$1)   &   100   &   $-$7.942 511 836 706($+$0)   \\
\hline
\end{longtable}

\newpage

\footnotesize
\begin{longtable}{@{\extracolsep\fill}rrrrrr@{}}
\caption{Magnetic shielding constants $\sigma$ (in units of $\alpha^2$) of Dirac one-electron atoms in the excited state 3d$_{3/2}$ ($\mu=\pm 5/2$). The number in brackets is the power of 10 by which the corresponding entry is to be multiplied. The value of the inverse of the fine-structure constant used was $\alpha^{-1}=137.035 \: 999 \: 139$ (from CODATA 2014).}
\label{tab:3d_5-2_mu_5-2}
$Z$ & \multicolumn{1}{c}{$\sigma[\alpha^2]$} & $Z$ & \multicolumn{1}{c}{$\sigma[\alpha^2]$} & $Z$ & \multicolumn{1}{c}{$\sigma[\alpha^2]$} \\
\hline\\
\endfirsthead
\caption[]{(continued)}
$Z$ & \multicolumn{1}{c}{$\sigma[\alpha^2]$} & $Z$ & \multicolumn{1}{c}{$\sigma[\alpha^2]$} & $Z$ & \multicolumn{1}{c}{$\sigma[\alpha^2]$} \\
\hline\\
\endhead
1    &   4.761 962 866 692($-$2)   &   47   &   2.299 299 490 566($+$0)   &   93    &   4.923 743 081 357($+$0)   \\
2    &   9.524 274 371 172($-$2)   &   48   &   2.350 949 746 169($+$0)   &   94    &   4.988 165 533 544($+$0)   \\
3    &   1.428 728 319 655($-$1)   &   49   &   2.402 775 961 817($+$0)   &   95    &   5.052 986 626 024($+$0)   \\
4    &   1.905 133 811 660($-$1)   &   50   &   2.454 782 185 846($+$0)   &   96    &   5.118 212 332 840($+$0)   \\
5    &   2.381 678 804 110($-$1)   &   51   &   2.506 972 491 314($+$0)   &   97    &   5.183 848 692 567($+$0)   \\
6    &   2.858 398 206 127($-$1)   &   52   &   2.559 350 976 615($+$0)   &   98    &   5.249 901 809 561($+$0)   \\
7    &   3.335 326 949 516($-$1)   &   53   &   2.611 921 766 103($+$0)   &   99    &   5.316 377 855 222($+$0)   \\
8    &   3.812 499 993 322($-$1)   &   54   &   2.664 689 010 726($+$0)   &   100   &   5.383 283 069 291($+$0)   \\
\hline
\end{longtable}

\newpage

\footnotesize
\begin{longtable}{@{\extracolsep\fill}rllllll@{}}
\caption{Comparison of present values (the upper entries) of the magnetic shielding constants $\sigma$ (in units of $\alpha^2$) for selected Dirac one-electron atoms in the states with zero radial quantum number, i.e: 1s$_{1/2}$, 2p$_{3/2}$, and 3d$_{5/2}$, with those reported by Moore \cite{Moor99} (the lower entries). All the numbers were obtained using $\alpha^{-1}=137.035 \: 989 \: 5$ (from CODATA 1986). The present results were computed from the analytical formula given by Eqs.\ (\ref{eq:1})--(\ref{eq:4}).}
\label{tab:comparison}
$Z$ & \multicolumn{1}{c}{$1s_{1/2}$} &  \multicolumn{1}{c}{$2p_{3/2}$ ($\mu=\pm 1/2$)} &  \multicolumn{1}{c}{$2p_{3/2}$ ($\mu=\pm 3/2$)} & \multicolumn{1}{c}{$3d_{5/2}$ ($\mu=\pm 1/2$)} & \multicolumn{1}{c}{$3d_{5/2}$ ($\mu=\pm 3/2$)} & \multicolumn{1}{c}{$3d_{5/2}$ ($\mu=\pm 5/2$)} \\
\hline\\
\endfirsthead
\caption[]{(continued)}
$Z$ & \multicolumn{1}{c}{$\sigma[\alpha^2]$} & $Z$ & \multicolumn{1}{c}{$\sigma[\alpha^2]$} & $Z$ & \multicolumn{1}{c}{$\sigma[\alpha^2]$} \\
\hline\\
\endhead
1  & 0.333 381 164 739 8 & $-$2781.968 506 04   & 0.100 002 853 429 7 & $-$1802.736 887 360  & $-$1201.808 718 363  & 0.047 619 628 667    \\
   & 0.333 381 19        & $-$2773.934 655 74   & 0.100 002 86        & $-$1789.538 468 54   & $-$1193.009 772 48   & 0.047 619 63     \\*[2ex] 
2  & 0.667 049 409 010 6 & $-$1390.856 472 495  & 0.200 022 828 498 3 & $-$901.317 586 311 9 & $-$600.846 643 293 3 & 0.095 242 743 712 37 \\
   & 0.667 049 43        & $-$1389.956 356 59   & 0.200 022 85        & $-$899.218 783 23    & $-$599.447 441 23    & 0.095 242 75     \\*[2ex]
3  & 1.001 292 268 112   & $-$927.095 665 082 3 & 0.300 077 052 149 9 & $-$600.821 882 288   & $-$400.500 297 248   & 0.142 872 831 967 7  \\
   & 1.001 292 32        & $-$927.202 616 7     & 0.300 077 04        & $-$600.433 022 1     & $-$400.241 063 86    & 0.142 872 84     \\*[2ex]
4  & 1.336 398 190 326   & $-$695.172 658 705 4 & 0.400 182 661 942   & $-$450.557 077 069 8 & $-$300.307 880 252 8 & 0.190 513 381 171 2  \\
   & 1.336 398 28        & $-$695.139 330 63    & 0.400 182 67        & $-$450.648 271 81    & $-$300.368 676 75    & 0.190 513 38     \\
\hline
\end{longtable}


\footnotesize
\begin{longtable}{@{\extracolsep\fill}rllll@{}}
\caption{Comparison of present values (the upper entries) of the magnetic shielding constants $\sigma$ (in units of $10^{-6}$) for selected Dirac one-electron atoms in the ground state (1s$_{1/2}$) and the first excited state, i.e: 2p$_{1/2}$ and 2p$_{3/2}$, with those obtained by Pyper and Zhang \cite{Pype99} (the lower entries). All the numbers were obtained using $\alpha^{-1}=137.035 \: 989 \: 5$ (from CODATA 1986). The present results were computed from the analytical formula given by Eqs.\ (\ref{eq:1})--(\ref{eq:4}).}
\label{tab:comparison2}
$Z$ &  \multicolumn{1}{c}{$1s_{1/2}$} &  \multicolumn{1}{c}{$2p_{1/2}$} &  \multicolumn{1}{c}{$2p_{3/2}$ ($\mu=\pm 1/2$)} & \multicolumn{1}{c}{$2p_{3/2}$ ($\mu=\pm 3/2$)}  \\
\hline\\
\endfirsthead
\caption[]{(continued)}
$Z$ & \multicolumn{1}{c}{$\sigma[\alpha^2]$} & $Z$ & \multicolumn{1}{c}{$\sigma[\alpha^2]$} & $Z$  \\
\hline\\
\endhead
1   & 17.753 001 077 21  & 148151.599 774 7    & $-$148143.611 901 5  & 5.325 288 145 917    \\
    & 17.8               & 148151.6            & $-$148143.6          & 5.3                  \\*[2ex]
11  & 198.677 736 235 3  & 13506.180 451 56    & $-$13418.088 958 59  & 58.779 118 929 66    \\
    & 198.7              & 13506.2             & $-$13418.1           & 58.8                 \\*[2ex]
19  & 355.241 951 732 8  & 7863.904 198 617 3  & $-$7710.948 135 58   & 102.225 649 690 6    \\
    & 355.23             & \phantom{666666}--- & \phantom{666666}---  & \phantom{666666}---  \\*[2ex]
37  & 801.389 998 548 4  & 4140.214 959 702    & $-$3835.155 688 112  & 204.893 238 544 4    \\
    & 801.4              & 4140.2              & $-$3835.2            & 204.9                \\
\hline
\end{longtable}


\section*{References}

\begin{thebibliography}{99}
\bibitem{Dalg62}
	 A. Dalgarno,
	 Advances in Physics 11 (1962) 281.
\bibitem{Kolb82}
	 D. Kolb, W. R. Johnson, P. Shorer,
	 Physical Review A 26 (1982) 19. 
\bibitem{John83}
	 W. R. Johnson, D. Kolb, K. N. Huang,
	 Atomic Data and Nuclear Data Tables  28 (1983) 333.
\bibitem{Hegs69}
	 R. A. Hegstrom,
	 Physical Review A 184 (1969) 17.
\bibitem{Szmy97}
	 R. Szmytkowski,
   Journal of Physics B 30 (1997) 825  
   [erratum: Journal of Physics B 30 (1997) 2747;
   addendum: arXiv:physics/9902050].
\bibitem{Stef12}
   P. Stefa{\'n}ska, R. Szmytkowski,
   International Journal of Quantum Chemistry 112 (2012) 1363. 
\bibitem{Szmy02b}
   R. Szmytkowski,
   Journal of Physics B 35 (2002) 1379. 
\bibitem{Szmy04}
   R. Szmytkowski, K. Mielewczyk,
   Journal of Physics B 37 (2004) 3961. 
\bibitem{Szmy02a}
   R. Szmytkowski,
   Physical Review A 65 (2002) 012503. 
\bibitem{Miel06}
   K. Mielewczyk, R. Szmytkowski,
   Physical Review A 73 (2006) 022511  
   [erratum:  Physical Review A 73 (2006) 039908].
\bibitem{Szmy11}
   R. Szmytkowski, P. Stefa{\'n}ska,
   e-print arXiv:1102.1811.
\bibitem{Szmy12}
   R. Szmytkowski, P. Stefa{\'n}ska,
   Physical Review A 85 (2012) 042502. 
\bibitem{Szmy14}
   R. Szmytkowski, P. Stefa{\'n}ska, 
   Physical Review A 89 (2014) 012501.
\bibitem{Szmy16}
	 R. Szmytkowski, G. {\L}ukasik,
	 Physical Review A 93 (2016) 062502.
\bibitem{Szmy16a}
	 R. Szmytkowski, G. {\L}ukasik,
	 Atomic Data and Nuclear Data Tables 111-112 (2016) 41.
\bibitem{Stef16}
	 P. Stefa{\'n}ska,
	 Physical Review A 94 (2016) 012508.
\bibitem{Stef15}
	 P. Stefa{\'n}ska, 
	 Physical Review A 92 (2015) 032504.
\bibitem{Stef16a}
	 P. Stefa{\'n}ska, 
	 Physical Review A 93 (2016) 022504.
\bibitem{Stef16b}
	 P. Stefa{\'n}ska, 
	 Atomic Data and Nuclear Data Tables  108 (2016) 193.
\bibitem{Stef17}
	 P. Stefa{\'n}ska, 
	 Atomic Data and Nuclear Data Tables  113 (2017) 316.
\bibitem{Mohr14}
	 P. J. Mohr, D. B. Newell, B. N. Taylor, 
	 CODATA Recommended Values of the Fundamental Physical Constants: 2014, 
	 arXiv:abs/1507.07956.
\bibitem{Moor99}
	 E. A. Moore,
	 Molecular Physics 97 (1999) 375.
\bibitem{Pype99}
	 N. C. Pyper, Z. C. Zhang,
	 Molecular Physics 97 (1999) 391.
\bibitem{Cohe87}
	E. R. Cohen, B. N. Taylor, 
	Reviews of Modern Physics 59 (1987) 1121.
\bibitem{Feio69}
	 F. D. Feiock, W. R. Johnson,
	 Physical Review A 187 (1969) 39.
\end{thebibliography}
\end{document}